\documentclass[letterpaper]{JHEP3}
\usepackage[dvips]{epsfig}
\usepackage{epsfig}
\usepackage{graphicx}

\title{Notes on Euclidean de Sitter space}
\author{V. Suneeta\\
Theoretische Physik, Ludwig-Maximilians Universit\"{a}t, \\
Theresienstrasse 37,
D-80333, M\"{u}nchen, Germany\\ E-mail: \email{suneeta@theorie.physik.uni-muenchen.de}} 

\abstract{
We discuss issues relating to the topology of Euclidean de Sitter space. We
show that in $(2+1)$ dimensions, the Euclidean continuation of the `causal
diamond', i.e the region of spacetime accessible to a timelike observer is
a three-hemisphere. 
However, when de Sitter entropy is computed in a `stretched horizon' picture, 
then we argue that the correct Euclidean topology is a solid torus.  
The solid
torus shrinks and degenerates into a three-hemisphere as one goes from the
`stretched horizon' to the horizon, giving the Euclidean continuation of
the causal diamond. 
We finally 
comment on generalisation of these results to higher dimensions.} 
\begin{document}

\section{Introduction}
De Sitter space has been of recent interest, mainly due to the discovery that
the universe has a small positive cosmological constant. De Sitter space
could thus be the natural limiting spacetime for the universe. Also, as it
is the most symmetric spacetime with a positive cosmological costant, it
is natural to investigate quantum gravity on de Sitter. This has 
already been attempted in several recent approaches \cite{banados, gks1, strominger, no, klemm}.
Another interesting feature of de Sitter space is the presence of a horizon
for a timelike observer. Associated with this horizon is an entropy,
similar to black holes. A microscopic description of this entropy may lead
to a better understanding of entropy associated with cosmological horizons
in more general contexts.

The issue of the microscopic origin of entropy for de Sitter space has
been dealt with in many different Lorentzian and Euclidean approaches,
particularly in $(2+1)$ dimensions \cite{banados, gks1, ms, flin, park, vb, myung, kabat}. 
While most of the computations
focus on the part of de Sitter space accessible to the timelike observer,
computations based on the recently proposed dS/CFT correspondence describe
de Sitter entropy in terms of degrees of freedom of a CFT at past/future 
infinity. In this letter, we discuss aspects related to the former approach,
namely, we take the view that the degrees of freedom corresponding to the
entropy are associated with information loss across the horizon of the timelike
observer. Therefore, it is the part of de Sitter spacetime accessible to
the timelike observer that is physically relevant (i.e the {\em causal
diamond}, a point of view emphasised in \cite{bousso}).

In computing the entropy in a Euclidean approach, this view must be reflected.
That is, we must look for the Euclidean continuation of the part of
de Sitter spacetime accessible to the timelike observer. This Euclidean 
continuation, and its topology are tricky issues - as we show in
the case of $(2+1)$-d de Sitter spacetime. The Euclidean continuation
of the part of spacetime seen by a timelike observer is {\em not} a
three-sphere, even though the metric continues to the three-sphere
metric. It is in fact, a three-{\em hemisphere}. However, there are more
subtleties - entropy of de Sitter space arises in the Euclidean 
approach in a `stretched horizon' picture which has been used earlier 
to explain black hole entropy \cite{carlip}. This makes the relevant 
{\em topology} not a three-hemisphere but a solid torus! The `stretched
horizon' picture is necessitated whenever boundary conditions are to be imposed 
at the horizon. In the Euclidean continuation, the horizon is a 
degenerate surface, and the boundary conditions must be imposed at the 
stretched horizon.

All these above
results are shown for $(2+1)$-d de Sitter spacetime. We also comment finally about 
generalisations of these statements to higher dimensions.

\section{Three dimensional Euclidean de Sitter}
As is well-known, de Sitter spacetime can be thought of as a hyperboloid
embedded in Minkowski spacetime of one dimension higher. 
\FIGURE{
\includegraphics{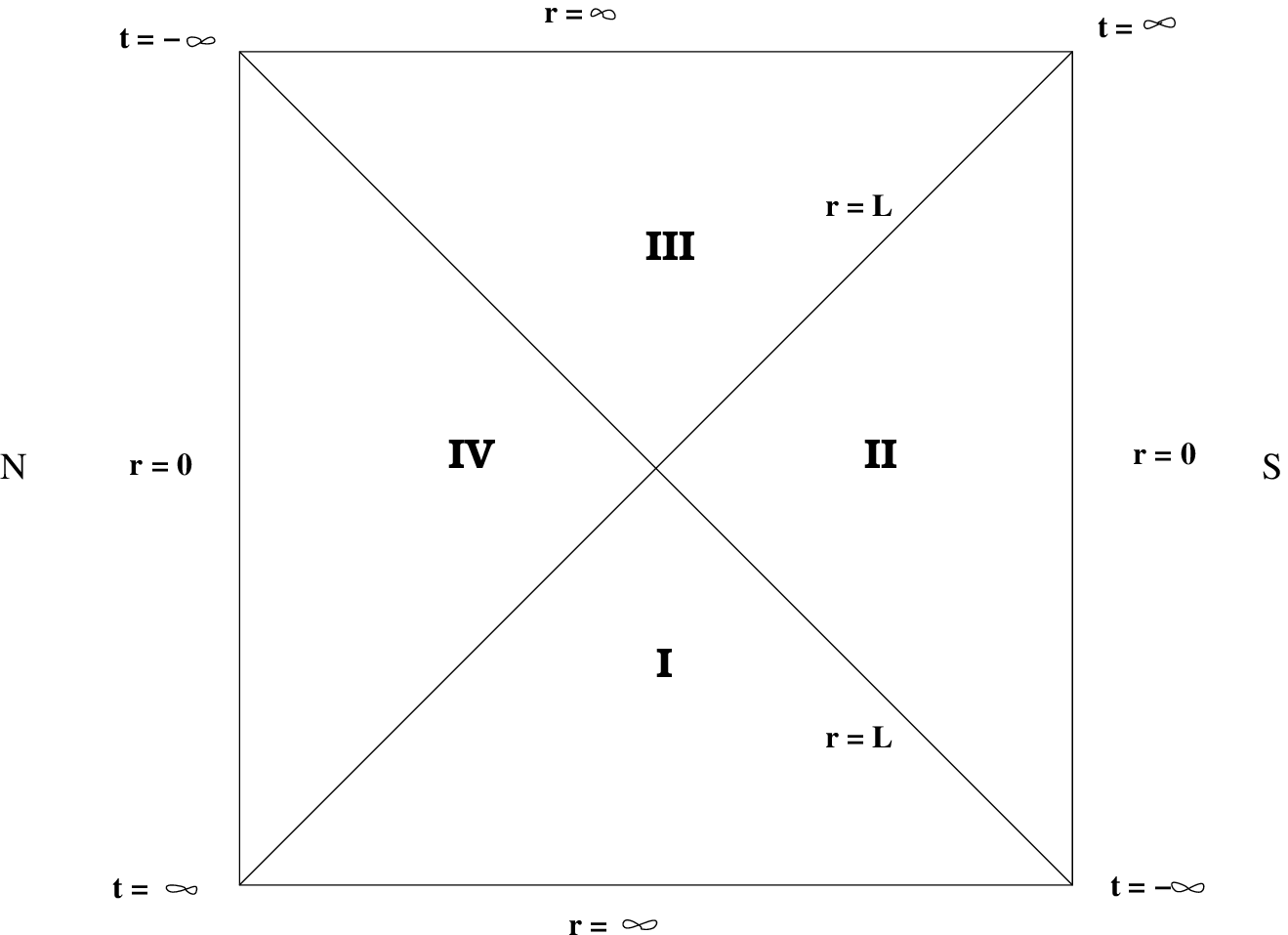}
\caption{}
}
The Penrose diagram of de Sitter space is given in Figure 1. A metric
that covers all of $D$ dimensional
de Sitter space with cosmological constant $\Lambda$,
called the global metric is

\begin{eqnarray}
ds^2 = - d\tau^2 + L^2 \cosh^2 (\tau/L) d\Omega_{D-1}^2
\label{global}
\end{eqnarray}

Here, $\Lambda = \frac{(D-1)(D-2)}{2 L^2}$. 
The topology of global de Sitter space is thus $S^{D-1} \otimes R$.
In these coordinates, however, there is no horizon. The metric is also 
time-dependent. One can describe the portion of de Sitter space visible
to a timelike observer (either patch II or IV in the Penrose
diagram with the observer at $r=0$) by the metric

\begin{eqnarray}
ds^2 = - V(r) dt^2 + V(r)^{-1} dr^2 + r^2 d\Omega_{D-2}^2
\label{static} 
\end{eqnarray}

where 
\begin{eqnarray}
V(r) = 1 - \frac{r^2}{L^2}
\end{eqnarray}

In these coordinates, the cosmological horizon is manifest, and lies at 
$r = L$. Thus, for a timelike observer, $0 \leq r \leq L$. 
The topology of the part of spacetime within this horizon 
is $B_{D-1} \otimes R$. Here $B_{D-1}$
refers topologically to a solid ball in $D-1$ dimensions, with a boundary
which is $S^{D-2}$.
Rewriting the metric (\ref{static}) with the change of coordinates
$r = L \sin \Theta$, we get
\begin{eqnarray}
ds^2 = - \cos^2 \Theta dt^2 + L^2 d\Theta^2 + L^2 d\Omega_{D-2}^2
\label{static2}
\end{eqnarray}
where the region within the horizon is given by
$0 \leq \Theta \leq \pi/2$ and the horizon is at $\Theta = \pi/2$. Equal 
time surfaces in these coordinates look like $(D-1)$-hemispheres. There is 
no contradiction with (\ref{static}), as a solid ball in $D$
dimensions is topologically
equivalent to a $D$-hemisphere. For example, a solid ball in two dimensions 
is a {\em disc} - which is equivalent topologically to a two-hemisphere.

We now look at the Euclidean continuations of (\ref{global}) and
(\ref{static}). We specialize in all the following discussions to the
case of $(2+1)$-d de Sitter spacetime.
Let us first study the Euclidean continuation of the global de Sitter 
metric (\ref{global}). 

The Euclidean continuation of the global
metric (\ref{global}) for three dimensional de Sitter space
is obtained by a Wick rotation of the time
coordinate $\tau_{E} = i\tau$. The period of $\tau_{E}$ is 
$\beta = 2\pi L$; obtained from the condition that the metric be regular
everywhere. To avoid rescaling and defining too many new coordinates, let us 
set $L=1$.
The metric is
\begin{eqnarray}
ds^2 = d\tau_{E}^{2} + \cos^{2} \tau_{E} (d\theta_{1}^2 
+ \sin^{2} \theta_{1} d\theta_{2}^{2}) 
\label{eucglobal}
\end{eqnarray}
where $\theta_{1}$ and $\theta_{2}$ are angles parametrising the 
two-sphere equal time surfaces of the Lorentzian global metric.
The metric (\ref{eucglobal}) is clearly the metric on a three-sphere, 
and therefore the topology of three dimensional global Euclidean
de Sitter space is $S^{3}$. The correct range of coordinates that 
cover the three-sphere completely are the usual 
$0 \leq \theta_{1} \leq \pi$, $0 \leq \theta_{2} \leq 2\pi$; and now
$-\pi/2 \leq \tau_{E} \leq \pi/2$.

One can also consider the Euclidean continuation of the metric
on the static patch (\ref{static}).
This is obtained by taking $t_{E} = it$.
The metric is
\begin{eqnarray}
ds^2~=~  (1 - r^2)~dt_{E}^{2} + (1 - r^2)^{-1}~dr^2 + r^2~d\phi^{2}
\label{eucstatic}
\end{eqnarray}
In addition, we make 
the Euclidean time periodic with its period
$\beta = 2\pi$ (as $L=1$).

We can make a coordinate change from the static
coordinates $(t_{E}, r, \phi)$ 
to the global coordinates $(\tau_{E}, \theta_{1}, \theta_{2})$  
through the transformations : 
\begin{eqnarray}
\phi~~&=&~~\theta_{2} \nonumber \\
r~~&=&~~\cos \tau_{E}~~\sin \theta_{1} \nonumber \\
\tan t_{E} ~~&=&~~
\frac{1}{\cos \theta_{1}}~~\tan \tau_{E} 
\label{st-gl}
\end{eqnarray}
These transformations cast the metric (\ref{eucstatic}) into the 
three-sphere metric (\ref{eucglobal}). However, when we describe de Sitter
space using the static patch metric (\ref{static}), we are interested in the
spacetime accessible to the timelike observer at $r=0$ which is either
patch II or IV, {\em not both}. The Euclidean continuation relevant to this
physical situation should reflect this fact. This should be naturally seen
in the range of the above coordinates - the relevant ranges are those for which
only one of the patches II or IV is described. 

In fact, the Euclidean continuation describing only one static patch is not a
sphere, but a hemisphere : the coordinate $\theta_{1}$ must only take half
the usual range required to cover the three-sphere. In order to see this,
we consider point masses in de Sitter space. In three dimensions, these
solutions have been well-studied by t'Hooft, Deser and Jackiw
for flat space in \cite{thooft}
and for non-zero cosmological constant by Deser and Jackiw in \cite{deser}.
The point mass geometries are simply wedges of de Sitter space with a wedge
proportional to the mass. Also, for the case of positive cosmological constant, there
is no {\em one-particle} solution and the lowest number of point masses one can have
is two. We reproduce below the form of the solution. 
In the original coordinates of Deser and Jackiw \cite{deser}, the Lorentzian static metric solution 
corresponding to point masses can be parametrised as
\begin{eqnarray}
- g_{00} = N^2(\bf r) \nonumber \\
  g_{0i} = 0 \nonumber \\
  g_{ij} = \psi(\bf r)
\label{ansatz}
\end{eqnarray}

The two-space can be expressed in terms of complex coordinates $z, \bar z$, where
$z = x + iy = R e^{i\chi}$.

The interval in two-space (at constant time) is
\begin{eqnarray}
dl^2 = \psi dz d\bar z
\end{eqnarray}

Following Deser and Jackiw, one can define the function $V(z)$ such that
\begin{eqnarray}
2 \psi^{-1} \partial_{\bar z} N = \Lambda V(z)
\end{eqnarray}

and a new variable $\zeta$
\begin{eqnarray}
\zeta = \frac{1}{2} \left[ \int^{z} \frac{dw}{V(w)} + \int^{\bar z} \frac{d\bar w}{V(\bar w)} \right ]
\label{zetadef}
\end{eqnarray}

Then, the form of the solution is  
\begin{eqnarray}
\psi = \frac{\epsilon}{\Lambda V(z) V^{*}(\bar z) \cosh^2 \sqrt{\epsilon}(\zeta - \zeta_{0})} \nonumber \\
N = \sqrt{\epsilon} \tanh \sqrt{\epsilon}(\zeta - \zeta_{0})
\label{soln}
\end{eqnarray}

Here, $\epsilon$ and $\zeta_{0}$ are constants. $\epsilon$ is restricted to be positive.
On performing the transformations 
\begin{eqnarray}
\sin \Theta = \frac{1}{\cosh \sqrt{\epsilon}(\zeta - \zeta_{0})} \nonumber \\
\phi = \frac{\sqrt{\epsilon}}{2i} \left [\int^{z} \frac{dw}{V(w)} - \int^{\bar z}
\frac{d\bar w}{V^{*}(\bar z)} \right ]
\label{transf}
\end{eqnarray}

the metric (\ref{ansatz}) then reduces to the static form of the de Sitter metric
(\ref{static2}) - except that the ranges for the coordinates will have to
be worked out carefully. In general, they will be different from the usual
ranges for de Sitter space, reflecting the presence of point masses in the 
spacetime.

Now, solutions corresponding to various distributions of point masses can be realised
by fixing a form for the function $V(z)$ - the strengths and locations of the 
zeroes and poles of $V$ then determine the masses and their locations. In the simplest
case,
\begin{eqnarray}
V = c^{-1} z
\label{v}
\end{eqnarray}
Then, also using the single-valuedness of $\chi$ and $\zeta$, we get $\zeta = c \ln R$.
We can rewrite $\zeta_{0} = c \ln R_{0}$.
Then,
\begin{eqnarray}
\psi = \frac{\epsilon c^2}{\Lambda R^2 \cosh^2 \sqrt{\epsilon} c \ln R/R_{0}}, \nonumber \\
N = \sqrt{\epsilon} \tanh \sqrt{\epsilon} c \ln R/R_{0} 
\label{twopart}
\end{eqnarray}

For (\ref{v}), we can perform the transformation (\ref{transf}) to the de Sitter
static coordinates and see the range and behaviour of $\Theta$ and $\phi$. We 
obtain
\begin{eqnarray}
\phi = \sqrt{\epsilon} c \chi \nonumber \\
\sin \Theta = \frac{2}{(R/R_{0})^{\sqrt{\epsilon}c} + (R/R_{0})^{-\sqrt{\epsilon}c}}
\label{twoparttrans}
\end{eqnarray}

Observing the form of the Deser-Jackiw solution (\ref{twopart}) and the 
transformation to the static coordinates $\phi$ and $\Theta$, we note the
following :

1) The periodicity of $\phi$ is not $2\pi$, but $2\pi \sqrt{\epsilon} c$
- to be interpreted as a wedge at the location of the point mass with
defect angle at the wedge  proportional to $(1 - \sqrt{\epsilon} c)$. 

2) There is however {\em no single} point mass solution! This is in
fact a two-particle solution.
As the coordinate $R$ goes over its full range from $0$ to $\infty$, 
$\Theta$ goes from $0$ to $\pi$. 
But at $\Theta = 0$ and $\Theta = \pi$, locally, we have a wedge - due to the 
non-standard periodicity of $\phi$. Thus constant time slices are spheres with
a wedge removed - corresponding to a point mass {\em each}
at the north and south poles making this a two-particle solution. 
In terms of the coordinate $r$ in the static metric (\ref{static}), $r$ goes from
$0$ to $1$ (horizon) and back to $0$ for this range of $\Theta$. Thus, the region around
each mass (the hemisphere) can be described by the coordinate $r$ and the mass is at
$r=0$. This is the region II or IV in the Penrose diagram in Figure 1 where the north(N)
and south(S) poles are indicated. This also makes clear that each timelike
observer can see only {\em one} mass - the region with the `mirror' mass
is in a causally disconnected part of spacetime.

Since this is a static solution, Euclideanisation is simply $t_{E} = it$, and $t_{E}$
is periodic. Thus the Euclidean continuation of the above solution is two point masses
at the antipodes of a {\em three}-sphere. But since 
a timelike observer can see only one of the two masses, in the Euclidean
continuation of the region {\em accessible} to this observer, we should not consider the full three-sphere, but the three-hemisphere.

This naturally remains true when the `strength' of the point mass 
- i.e, the defect angle is taken to zero, and we
have empty de Sitter space \footnote {We note here that the notion of mass in de Sitter space is not well-defined due to the absence of a globally timelike Killing vector.}. Thus, in (\ref{st-gl}), this implies that the range of
coordinates should be such that they cover the hemisphere - and therefore,
$0 \leq \theta_{1} \leq \pi/2$. 

\section{De Sitter entropy in three dimensions}
We have shown above that the static patch metric describing the part
of spacetime seen by the timelike observer has a Euclidean continuation
which is the metric on a three-sphere: however, it covers only the
three-hemisphere due to the correct range of coordinates. 
In this section, we examine the question of what is the
relevant manifold for describing de Sitter {\em entropy} in a Euclidean
computation. Surprisingly, the answer is not a three-hemisphere, but a solid 
torus. Each static patch (II or IV in Figure 1) has a topology $D_{2} \otimes R$
where $R$ indicates the time direction. The boundary of the disc $D_{2}$ is 
the horizon. Now, since time is made periodic in the Euclidean continuation,
for any {\em fixed} $r < 1$ in (\ref{eucstatic}), the topology of the Euclidean
surface is a torus. If one looks at the Euclidean manifold for all values of
$r < 1$, this 
is a solid torus - with $\phi$ parametrising the contractible cycle and the
Euclidean time $t_{E}$ parametrising the non-contractible cycle. However,
with respect to the metric (\ref{eucstatic}), as $r$ varies from $0$ to 
$1$, the solid torus {\em shrinks} in size, as indicated in Figure 2.
More precisely, the contractible cycle grows and the non-contractible cycle 
shrinks. At the horizon $r=1$, the contractible cycle has a maximum radius
of $1$, and the non-contractible cycle degenerates to a point. Thus we have
a topology change from a solid torus characterised by a non-contractible cycle
to a solid ball. But since a solid ball is topologically the same as the
three-hemisphere, we have recovered the three-hemisphere picture of the
previous section. However, we are left with the following question: Which
is the topology that correctly describes the entropy degrees of freedom in 
the Euclidean picture? Must we consider the three-hemisphere or the solid 
torus?
\FIGURE{
\includegraphics{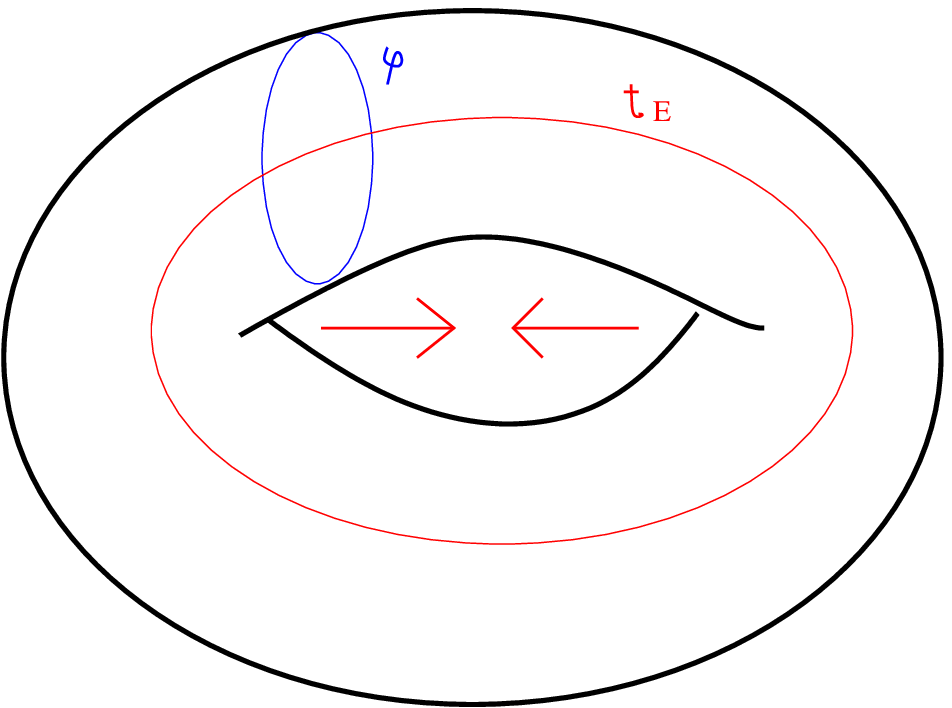}
\caption{}
}
Before answering this question, it is instructive to look at :

1) The analogous situation for black holes, i.e what is the relevant 
Euclidean manifold that correctly describes black hole entropy?

2) Computations of de Sitter entropy in Lorentzian/Euclidean approaches.

We restrict ourselves to a study of entropy computations of the BTZ black
hole \cite{btz} also in $(2+1)$ dimensions but with a {\em negative} cosmological
constant - as there are many similarities with our case. The entropy 
computations exploit the connection between gravity in three dimensions 
and Chern-Simons theory \cite{at} and the relevant degrees of freedom are
degrees of freedom of a Chern-Simons theory on a manifold 
with boundary. It is well-known that standard Euclidean continuations of
black hole spacetimes describe the region from the horizon to infinity and
the horizon is a degenerate surface in the Euclidean manifold. It has been
shown in \cite{cat} that the Euclidean continuation of the BTZ black hole
is a solid torus. The horizon is a degenerate circle at the core of the
solid torus parametrised by an angular coordinate $\phi$ and situated at
$r=0$ in
Schwarzschild-like coordinates. The Euclidean
time is now the contractible cycle of this solid torus. A partition function
for this black hole in the Euclidean path integral approach has been derived 
in \cite{gks} using results from
Chern-Simons theory. It was shown before in \cite{carlip} that the leading 
contribution
to the entropy could be obtained by considering boundary degrees of freedom
of Chern-Simons theory, where the boundary is a `stretched horizon'. The 
horizon itself is a circle, and the stretched horizon is a torus tube 
surrounding this horizon, i.e at a small non-zero value of $r$. 
One then imposes 
invariance under residual diffeomorphisms (after imposition of boundary conditions) on
the boundary states at the stretched horizon. This yields the correct semi-classical entropy
on counting the states.

Now, looking at the analytic continuation of de Sitter from this perspective,
we see that the de Sitter static patch is like `the inside' of the black
hole, i.e the horizon is now the {\em outer}, rather than the {\em inner}
boundary of the physically accessible spacetime. In the BTZ case, the
Euclidean topology is a solid torus. Euclidean time is a contractible cycle
of the solid torus that shrinks to zero at the horizon. The horizon itself is a degenerate
{\em circle} at the core of the solid torus. However, the {\em stretched} horizon 
at $r > 0$ is not a circle but a torus.
In the Euclidean de Sitter, we again have a solid torus topology - but Euclidean time
is now the non-contractible cycle.
We have a degenerate case at the horizon $r=1$, where  
the {\em non-contractible cycle} degenerates to a point - changing the
topology from solid torus to the hemisphere. However, the stretched
horizon is now at a value $r < 1$, and its topology is again a torus as before. 
Considering the 
region from $r=0$ to the {\em stretched} horizon, the topology is that of a 
solid torus. Thus
if the degrees of freedom corresponding to the entropy reside on the
stretched horizon, the three-dimensional topology of interest is a solid torus.

Let us examine the various computations of de Sitter entropy both in the
Lorentzian and Euclidean approaches. Maldacena and Strominger \cite{ms} have 
used the Chern-Simons formulation of three-dimensional gravity to describe the
entropy. They consider Chern-Simons theory on the Lorentzian static patch manifold
- which is $D_{2} \otimes R$. Then, boundary conditions are imposed on the fields at
the cylindrical boundary which is the horizon. These boundary conditions are motivated
by earlier work on BTZ black holes \cite{carlip1} and encode the fact that the boundary
is an apparent horizon. Then, a WZW conformal field theory is induced on this boundary. 
The physical states must obey the remnant of the Wheeler-de Witt equation, i.e invariance
under diffeomorphisms which preserve the boundary conditions. A simple counting of
states gives the correct entropy, and the states lie on the cylindrical boundary. If
an analogous computation were attempted in the Euclidean picture, it seems that the 
degrees of freedom on the cylindrical boundary would now be the degrees of freedom on the
torus obtained by compactifying time. Also, as we saw in the {\em Euclidean} BTZ entropy
computation, the imposition of the remnant of the Wheeler-de Witt equation will have to be 
done at the {\em stretched} horizon, as the horizon itself is a degenerate surface in 
the Euclidean picture. Considering the stretched horizon in the Euclidean de Sitter case implies
as mentioned before, that the relevant Euclidean manifold is a solid torus. 

There have been Euclidean computations using the solid torus topology. In \cite{banados},
it was claimed that the relevant space was an infinitesimal tube around the timelike
observer, which on Euclideanisation is a solid torus. More recently, a partition function
for de Sitter space was proposed \cite{gks1} using results from Chern-Simons theory on
a solid torus. From this computation, it is clear that the partition function is independent
of the location of the toral boundary. This computation gives the correct semi-classical
entropy and also predicts a first order correction to this entropy, which agrees with 
corrections obtained from other unrelated approaches, like the dS/CFT correspondence.

\section{Generalisation to higher dimensions}

The above discussion strongly points to the relevant Euclidean manifold for computing
$dS_{3}$ entropy being a solid torus. But we have argued this taking specific instances
in three dimensions, which use the connection between Chern-Simons theory and gravity.
A natural question to ask would be, how much of this generalises to higher dimensions?
The Euclideanisation of $n$-dimensional global $dS_{n}$ is an $n$-sphere - and by previous
arguments, the Euclidean continuation of the region relevant to each timelike observer is an $n$-hemisphere.
Likewise, the static patch has a topology $B_{n-1} \otimes R$ - where $B_{n-1}$ is an
$(n-1)$ dimensional solid ball, and $R$ denotes the time direction. The Euclidean
continuation of this, for $r < 1$ (horizon) is $B_{n-1} \otimes S^{1}$. Including
$r=1$ makes this manifold an $n$-hemisphere. Although many aspects of the discussions
in the previous sections are specific to three dimensions, it is true that in a 
Euclidean continuation of a static metric with horizon, the horizon becomes a 
degenerate surface. Thus in an entropy computation involving boundary conditions at
the horizon, in the Euclidean continuation, the boundary conditions must be imposed
at the {\em stretched} horizon. It is therefore
likely that the 
importance of the stretched horizon in Euclidean computations of black hole entropy
may be general, and valid in any dimension. This suggests that in any physical
situation involving the region of $dS_{n}$ seen by the timelike observer, 
particularly for understanding entropy, the correct
topology in the {\em Euclidean} picture is $B_{n-1} \otimes S^{1}$.
It would be interesting to explore the consequences of such an idea in cosmological 
situations.

\section*{\bf Acknowledgements.}
I would like to thank S. Carlip for useful comments on the draft of
this paper.
This work is supported by a fellowship from the 
Alexander von Humboldt Foundation.

\end{document}